\documentclass[prl,superscriptaddress,nobibnotes,twocolumn,showkeys,10pt]{revtex4-1} %
\usepackage{siunitx}
\usepackage{xcolor}
\usepackage{graphicx}
\usepackage[colorlinks=true,linkcolor=blue,citecolor=blue]{hyperref}
\usepackage{amsmath}
\usepackage{soul}

\begin{document}

\title{Controlling the non-linear emission of upconversion nanoparticles to enhance super-resolution imaging performance}

\author{Simone De Camillis}
\email{simone.decamillis@mq.edu.au}
\affiliation{ARC Centre of excellence for Nanoscale BioPhotonics (CNBP), Department of Physics and Astronomy, Macquarie University, NSW 2109, Australia}
\author{Peng Ren}
\affiliation{ARC Centre of excellence for Nanoscale BioPhotonics (CNBP), Department of Physics and Astronomy, Macquarie University, NSW 2109, Australia}
\affiliation{CNBP, School of Engineering and Physics, Macquarie University}
\author{Yueying Cao}
\affiliation{ARC Centre of excellence for Nanoscale BioPhotonics (CNBP), Department of Physics and Astronomy, Macquarie University, NSW 2109, Australia}
\affiliation{CNBP, Department of Molecular Sciences, Macquarie University}
\author{Martin Ploschner}
\affiliation{ARC Centre of excellence for Nanoscale BioPhotonics (CNBP), Department of Physics and Astronomy, Macquarie University, NSW 2109, Australia}
\affiliation{School of Information Technology and Electrical Engineering, The University of Queensland, QLD 4072, Australia}
\author{Denitza Denkova}
\affiliation{ARC Centre of excellence for Nanoscale BioPhotonics (CNBP), Department of Physics and Astronomy, Macquarie University, NSW 2109, Australia}
\affiliation{Institute for BioEngineering of Catalonia (IBEC), 08028 Barcelona, Spain}
\author{Xianlin Zheng}
\affiliation{ARC Centre of excellence for Nanoscale BioPhotonics (CNBP), Department of Physics and Astronomy, Macquarie University, NSW 2109, Australia}
\author{Yiqing Lu}
\email{yiqing.lu@mq.edu.au}
\affiliation{ARC Centre of excellence for Nanoscale BioPhotonics (CNBP), Department of Physics and Astronomy, Macquarie University, NSW 2109, Australia}
\affiliation{CNBP, School of Engineering and Physics, Macquarie University}
\author{James A. Piper}
\affiliation{ARC Centre of excellence for Nanoscale BioPhotonics (CNBP), Department of Physics and Astronomy, Macquarie University, NSW 2109, Australia}

\date{\today}

\begin{abstract}
\textbf{
Upconversion nanoparticles (UCNPs) exhibit unique optical properties such as photo-emission stability, large anti-Stokes shift, and long excited-state lifetimes, allowing significant advances in a broad range of applications from biomedical sensing to super-resolution microscopy. In recent years, progress on nanoparticle synthesis led to the development of many strategies for enhancing their upconversion luminescence, focused in particular on heavy doping of lanthanide ions and core-shell structures. In this article, we investigate the non-linear emission properties of fully Yb-based core-shell UCNPs and their impact on the super-resolution performance of stimulated excitation-depletion (STED) microscopy and super-linear excitation-emission (uSEE) microscopy. Controlling the power-dependent emission curve enables us to relax constraints on the doping concentrations and to reduce the excitation power required for accessing sub-diffraction regimes. We take advantage of this feature to implement multiplexed super-resolution imaging of a two-sample mixture.}
\end{abstract}


\maketitle

\section*{Introduction}

\hyphenation{au-to-fluo-resc-ence}
\hyphenation{fluo-resc-ence}
\hyphenation{prop-er-ties}
\hyphenation{per-for-mance}
\hyphenation{com-pa-ra-ble}
\hyphenation{lu-mi-nes-cence}
\hyphenation{sat-u-ra-tion}
\hyphenation{cor-re-spond-ing}
Lanthanide-based upconversion nanoparticles (UCNPs) efficiently convert absorbed near-infrared (NIR) light into higher photo-energy emission, enabling a large range of applications spanning biomedical imaging and sensing, drug-delivery systems, NIR vision extension, solar-cell technology, data security, and more \cite{Wu2015,Zhu2019,Balabhadra2017,Yang2020,Ma2019,Ambapuram2020}. 
Using UCNPs as luminescent bioprobes offers key advantages including extended photo-stability over several hours of laser exposure, excitation in the NIR spectral region resulting in low autofluorescence background and reduced degree of photo-toxicity for biological samples, and lifetime-based multiplexing capability \cite{Zheng2015,Wang2018,Fan2018}.

The most common UCNP structure comprises the crystal lattice NaYF$_{4}$ hosting the trivalent ions Yb$^{3+}$ and Tm$^{3+}$, which together represent a suitable sensitiser/activator pair. The Yb$^{3+}$ ion absorbs NIR ($970$-\SI{980}{\nano\metre}) radiation and efficiently transfers this energy in a step-wise manner to the Tm$^{3+}$ ion, which is responsible for the upconversion emission. 
The peculiar photo-physics of this class of nanoparticles has enabled significant advances in super-resolution microscopy. STimulated Excitation-Depletion (STED) microscopy \cite{Hell1994} for instance can be successfully applied to UCNPs, thanks to their efficient cross-relaxation mechanism from highly-excited states to intermediate energy levels within the Tm$^{3+}$ ions \cite{Liu2017,Zhan2017,Peng2019}. This phenomenon allows the blue emission to be substantially suppressed by efficient depletion of the intermediate energy levels via resonant stimulated emission. Imaging in combination with an annular-shaped depletion beam can therefore significantly improve the lateral resolution, even down to the nanoparticle's size. Optical investigation of the non-linear emission properties of UCNPs has also led to the discovery of an additional super-resolution technique \cite{Denkova2019,Caillat2013,Wang2016}, recently named upconversion Super-linear Excitation-Emission (uSEE) microscopy. With suitable concentrations of the lanthanide ion Tm$^{3+}$, the 455-nm emission of UCNPs shows strong non-linear dependence as function of the excitation laser intensity, reaching power indices of $6.2$. Such strong super-linearity results in improvement to the resolution down to about half of the diffraction limit in both the lateral and transversal directions, making uSEE microscopy a simpler, low-power super-resolution technique, not requiring complex purpose-build systems or any post-processing analysis.
The ability to resolve single UCNP probes in biological samples is important since this allows binding of individual nanoparticles to target structures to be visualised and quantified \cite{Denkova2019,Cao2020}.

Despite the unique advantages of UCNPs, their relatively moderate luminescence brightness represents a major challenge. Applications in STED nanoscopy in particular require a critical balance between high excitation laser intensities, which are potentially harmful for biological samples, and excessive acquisition or processing times. The uSEE technique is also affected, as the emission rate is limited to photon-count values of \SI{10}{cts\per\milli\second} \cite{Denkova2019}. 
Over the past decade, there have been major efforts directed at refining design and synthesis of lanthanide UCNPs aimed at enhancing their brightness \cite{Han2014}. This has been achieved, for example, by optimising the lanthanide-ion concentrations, and by growing active/inert shell structures \cite{Zheng2018,Liu2019,Homann2018}. Higher sensitiser (Yb$^{3+}$) concentrations provide a higher density of optical centres dedicated to the collection of photons and to the sustenance of the excitation process, leading to a stronger luminescence signal. At the same time, however, the reduced dopant separation facilitates ion-ion interactions and promotes long distance energy migration to quencher sites, therefore limiting the upconversion efficiency \cite{Chen2019}.
The strategy of overcoating the nanoparticle with shells of active and/or inactive materials has been shown effective in reducing the influence of quenching defects, as they are largely located on the surface of the particle \cite{Fischer2016,Tian2018}. For instance, the design of active sandwich structures has been adopted to physically separate activator and sensitiser in different layers, ensuring energy migration to the nanoparticle core \cite{Qiu2014} or preventing deleterious cross-relaxation back to the sensitiser ions \cite{Zhong2014}. On the other hand, the simple coating of a thin inert shell over a single active core has more recently been demonstrated to successfully boost the luminescence signal by one or even two orders of magnitude, specifically in a low-excitation power regime \cite{Tian2018,Shi2017}.
Finally, to further maximise the luminescence brightness several research groups have attempted the synthesis of lanthanide-based lattice, i.e.\ NaYbF$_{4}$:Ln (Ln: lanthanide), where the inert Y element is completely replaced by the Yb$^{3+}$ ions. Despite the possible limiting factor of concentration quenching \cite{Zheng2018}, significant improvements in brightness compared to the common Y-based UCNPs have been reported \cite {Shi2017,Shen2017,Tian2018,Cao2020}.

Given the above advances in UCNP design and synthesis, the question arises whether there are consequential improvements that can be made to super-resolution imaging performance. For example, changes in the Yb/Tm composition ratio are expected to affect excitation and depletion rates and the efficiency of cross-relaxation mechanisms, leading to variations in the non-linear emission characteristics of the excited nanoparticles.
Promising results have been reported for NaYF$_4$:Yb,Tm nanoparticles \cite{Ma2020}, where increasing the doping concentration of the Yb$^{3+}$ ions from $20$ to $60\%$ showed improvements in the emission slope at the single-nanoparticle level. However, power-dependent studies on fully Yb-based UCNPs are limited in literature, and to the best of our knowledge a comprehensive analysis of their emission properties and their effect on super-resolution nanoscopy of the particles have not been reported. 

In this paper, we report results of a systematic investigation of NaYbTmF$_4$ nanoparticles at different lanthanide ion ratios and over an extensive range of excitation powers. We demonstrate that these particles can efficiently sustain the depletion process for the STED microscopy, while also providing sufficient levels of super-linearity to enable uSEE microscopy over a wide selection of lanthanide concentrations. We observe that the emission curve of the single nanoparticle can be substantially shifted towards lower excitation powers by varying the lanthanide ion ratio, while the super-linear slope can be increased with the coating of an inert shell.
Further, optimising UCNP composition and structure enables state-of-the-art super-resolution imaging for both STED and uSEE microscopy at significantly lower excitation and depletion power densities that have been previously reported.
Additionally, we exploit this phenomenon to implement a multiplexing imaging method capable of recognising different UCNPs in a mixture with optimal super-resolution performance.

\section*{Results and discussion}

\begin{figure}[b]
  \centering
  \includegraphics{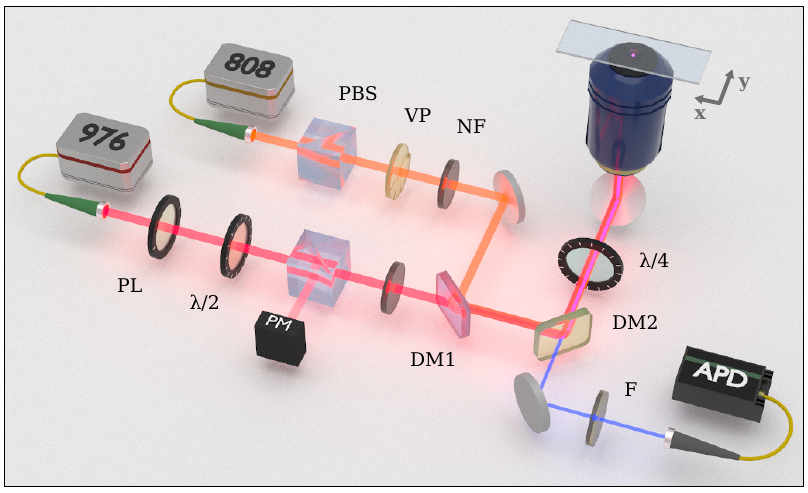}
  \caption{Schematic of the confocal microscope system, where PBS is a polarising beam splitter, VP a vortex plate, NF a neutral density filter, PL is a polariser, $\lambda/2$ a half-wave plate,  PM a power meter, DM1 and DM2 respectively a long-pass and short-pass dichroic mirror, $\lambda/4$ a quarter-wave plate, F a blue narrow-band filter, and APD an avalanche photo-diode.}
  \label{fgr:set-up}
\end{figure}

\begin{figure*}[!t]
  \centering
  \includegraphics{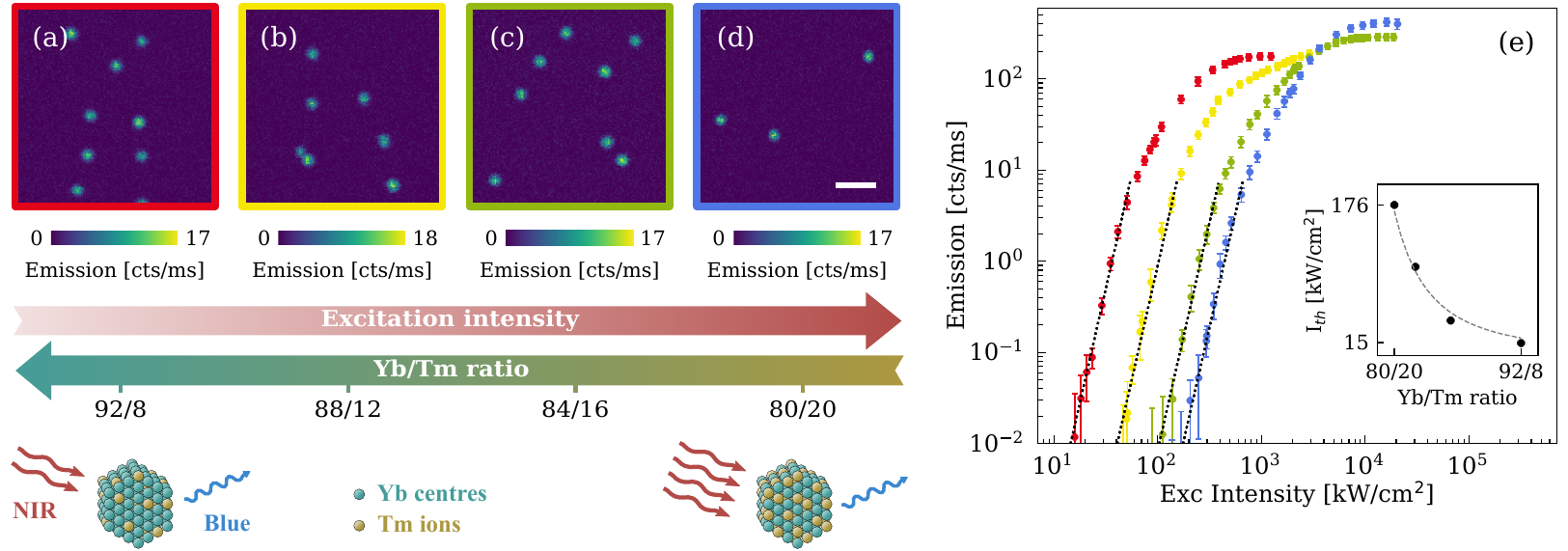}
  \caption{Imaging of lanthanide-based nanoparticles NaYb$_x$Tm$_{1-x}$F$_4$, where the Yb concentration $x = 0.92$, $0.88$, $0.84$, $0.8$. We refer to each sample with the Yb/Tm concentration ratio $\eta = 92/8$, $88/12$, $84/16$, and $80/20$, represented by the colour red, yellow, green, and blue, respectively. (a-d) Confocal $x$-$y$ images of the monodispersed UCNPs, where each sample was excited respectively at $I = 54$, $153$, $444$, and \SI{573}{\kilo\watt\per\centi\metre\squared}, to reach the same single-particle photo emission. Dwell time: \SI{3}{\milli\second}. Bar scale: \SI{1}{\micro\metre}. (e) Single-nanoparticle $455$-\SI{}{\nano\metre} emission as a function of the excitation peak intensity. The dotted lines illustrate a slope of 5. Inset: The luminescence signal overcomes the noise level of \SI{e-2}{cts\per\milli\second} at the intensity threshold $I_{\textrm{th}}$, which differs for each UCNP sample. $I_{\textrm{th}}$ is represented as function of the concentration ratio $\eta$.}
  \label{fgr:core-only}
\end{figure*}

A set of hexagonal NaYb$_x$Tm$_{1-x}$F$_4$ nanoparticles with Yb concentration of $x = 0.92$, $0.88$, $0.84$, and $0.8$ were synthesised following a thermal decomposition protocol. The synthesis was carefully realised by using a customised system for automated growth, ensuring a high degree of precision and reliability. Within each sample, particle size was observed to be homogeneous, with diameters $54$, $46$, $48$, and \SI{59}{\nano\metre}, respectively (see Supplementary Figure 1).
Note the concentrations of the sensitiser and the activator species are directly related to each other, so that increasing the former results in decreasing the latter. Therefore, we refer to the different nanoparticle compositions with the ratio $\eta =$Yb$_\% / $Tm$_\%$, representing the proportion between the Yb and Tm concentrations.

We performed the imaging and the optical characterisation by means of a custom-made dual-laser confocal system \cite{Denkova2019} schematically represented in Figure \ref{fgr:set-up} (full details in Experimental Methods). The UCNPs are excited by a $976$-\SI{}{\nano\metre} circularly-polarised laser beam, initially produced by a single-mode CW laser diode. 
Subsequently, a long-pass dichroic mirror allows an additional annular-shaped laser beam, centred at \SI{808}{\nano\metre}, to spatially co-propagate with the primary excitation beam. This secondary arm is unblocked only to perform STED microscopy. The excitation and depletion beams are delivered to the UCNP sample via a 100x microscope objective. The collected upconversion signal is isolated from the source component by a short-pass dichroic mirror, then filtered at the narrow wavelength range of \SI{446(13)}{\nano\metre}, and finally coupled into a multimode fibre for luminescence signal acquisition by an avalance photo diode.

We firstly acquired confocal images for each set of nanoparticles, as illustrated in Figure \ref{fgr:core-only}(a-d). A common luminescence yield of about \SI{17}{cts\per\milli\second} was reached by tuning the excitation intensity. Interestingly the same response was obtained under different excitation power densities, from \SI{573}{\kilo\watt\per\centi\metre\squared} in the case of the lowest value of concentration ratio $\eta = 80/20$, down to only \SI{54}{\kilo\watt\per\centi\metre\squared} for the highest value of $\eta = 92/8$. 
To investigate this phenomenon in more detail, we measured the luminescence of an isolated nanoparticle over a large range of excitation intensities. Results are reported in Figure \ref{fgr:core-only}(e) in a log-log graph to better evaluate their power dependence. 

All four samples show an initial strong slope at lower power, progressively reducing to a linear proportionality and eventually reaching a saturation regime. Generally, the slope $s$ of the emission curve is closely related to the multi-photon nature of the excitation process and, in the particular case of an upconversion phenomenon, it is equal to or smaller than the upconversion multi-photon number $N$ involved \cite{Pollnau2000}. Here, the blue luminescence centred at \SI{455}{\nano\metre} and assigned to the $^1$D$_2$ $\rightarrow$ $^3$F$_4$ transition has been often referred to as a five-photon process \cite{Chen2007,Wang2008}.
Our results are in good agreement with this assumption, as illustrated in Figure \ref{fgr:core-only}(e) by the slope $s = 5$ represented in dotted lines. 
It is important to note that energy redistribution mechanisms, such as excited state absorption and cross-relaxation between activator ions, are critically involved in the upconversion luminescence. Indeed, these processes, which are strongly influenced by the concentrations of the dopant ions, are believed to be responsible for reducing or increasing the emission slope of lanthanide-doped materials \cite{Pollnau2000,Wang2008,Bednarkiewicz2019}. Remarkably, despite the significant difference in the concentration ratio $\eta$, all four types of nanoparticles showed a comparable slope $s = 5$ at relatively low excitation intensity.

\begin{figure*}[t]
  \centering
  \includegraphics{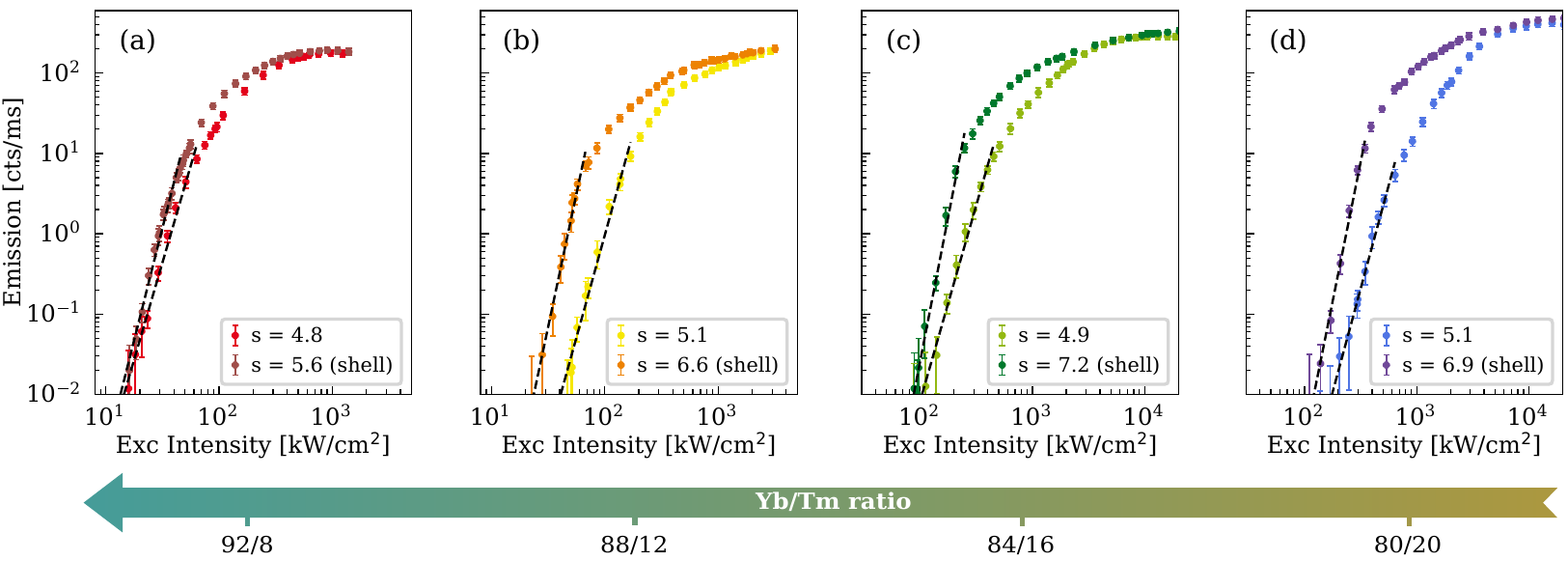}
  \caption{Comparison of the single-nanoparticle emission between core-only (in brighter colours) and core-shell (in darker colours) structures. The core-shell curves corresponding to $\eta = 92/8$, $88/12$, $84/16$, and $80/20$ are represented in dark red, orange, dark green, and purple, respectively. The brighter curves follow the color code as specified in Figure \ref{fgr:core-only}. The legend specifies the slope values obtained from the best fit (dashed line) of the emission curves at low laser intensities. The slope uncertainties are equal to $\pm 0.1$, with the exception of $\pm 0.2$ for the core-shell case of (c). The intensity threshold for each core-shell UCNP is equal to $13$, $23$, $87$, and \SI{121}{\kilo\watt\per\centi\metre\squared}, respectively.}
  \label{fgr:core-shell}
\end{figure*}

The notion that the emission curves in Figure \ref{fgr:core-only}(e) have a similar super-linear slope suggests that similar emission properties can be achieved at different excitation intensities by simply adjusting the concentration ratio. 
We quantified this shift by looking at the intensity threshold $I_{\textrm{th}}$ at which the linear best-fit of the luminescence signal overcomes the noise level of \SI{e-2}{cts\per\milli\second}. This is calculated for each sample and represented in the inset of Figure \ref{fgr:core-only}(e), showing an almost inverse proportionality to the square root of the concentration ratio $\eta$, within the range of parameters explored in this analysis. By increasing $\eta$ from $80/20$ to $92/8$, we were able to lower the power density of the excitation laser by more than an order of magnitude. 
This shift can be widely attributed to the larger concentration of Yb$^{3+}$ ions, as suggested by comparing the present results with the emission curve of the common Y-based UCNPs at the same doping level of the Tm$^{3+}$ ions (Supplementary Figure 2). Indeed, although higher Yb concentrations facilitate energy migration to surface quenchers, overall the fully Yb-doped nanoparticle with concentration ratio $\eta = 92/8$ (NaYb$_{0.92}$Tm$_{0.08}$F$_4$) shows a threefold decrease of the intensity threshold $I_{\textrm{th}}$ with respect to the corresponding 20\%-Yb-doped nanoparticle (NaY$_{0.72}$F$_4$:Yb$_{0.20}$,Tm$_{0.08}$) used in our previous work \cite{Denkova2019}.

At higher excitation intensity we note that the emission curves in Figure \ref{fgr:core-only}(e) reach a saturation region where the luminescence is almost constant as a function of laser power. The saturation level can be expected to vary according to the concentration of the lanthanide ions and their relative energy transfer efficiency. For an accurate comparison of the nanoparticle brightness at saturation the UCNP samples should have identical size, which is not the case here. We can nevertheless observe that the normalisation of the emission curves for the nanoparticle volume suggests a slight increase in the saturation yield with increasing concentration of Tm$^{3+}$ ions, while the slope values are effectively unchanged (Supplementary Figure 3).

To further improve the optical performance of our UCNPs, we coated the nanoparticles with a thin shell of NaYF$_4$. The morphological characterisation by TEM gives an average shell thickness of \SI{1.5}{\nano\metre} (Supplementary Figure 4). We observed a systematic increase of the emission slope for core-shell compared with core-only nanoparticles, as indicated by the data of Figure \ref{fgr:core-shell}. This is in agreement with results of previous studies of coating effects in high-Yb-doped UCNPs \cite{Ma2020}.
We attribute the slope improvement to the presence of the inert shell, which successfully reduces energy transfer from the Yb$^{3+}$ ions to the surface quenchers. A simplified energy-level model of the sensitiser/activator system confirms this hypothesis, estimating a slope change in the order of $20\%$ (Supplementary Note 1).
Each of the four core-shell UCNP samples examined herein demonstrated a non-linear slope value equal to or greater than $5.6$, with a maximum value of $7.2$ for $\eta = 84/16$.
The experimental results also suggest a trend in the slope improvement as function of the lanthanide concentrations, with less evident effects of the shell coating at higher values of the concentration ratio $\eta$. This behaviour can be rationalised by recalling that the $455$-nm emission is considered a five-photon process, meaning that at least five Yb$^{3+}$ ions are required to excite a Tm$^{3+}$ ion. Here, $\eta$ varies from $4$ (sample $80/20$) to $11.5$ (sample $92/8$), increasingly above the optimal Yb/Tm ratio. We can expect that at a certain threshold of this ratio, the Yb$^{3+}$ ions are more effective in exciting the Tm$^{3+}$ ions despite the presence of energy losses, making the shell coating improvement less evident (as in the case of sample $92/8$).
Shell coating has also shown the potential to further reduce the intensity threshold towards lower excitation powers. The different shift in intensity threshold is influenced by a certain degree of inhomogeneity in the shell thickness among nanoparticles. Although these variations have limited effect for $\eta = 92/8$, larger shifts were observed for $\eta = 80/20$ (see Supplementary Figure 5).
In addition, we emphasise that the shell coating has little effect at higher power densities, despite having a strong impact in the super-linear emission region. At high power densities, the effective excitation rate of the Tm centres well exceeds the non-radiative energy loss rate related to the surface quenching, thus limiting the maximum emission rate of the single core-shell UCNPs to the same values as the corresponding core-only nanoparticles.

\begin{figure}[!t]
  \centering
  \includegraphics{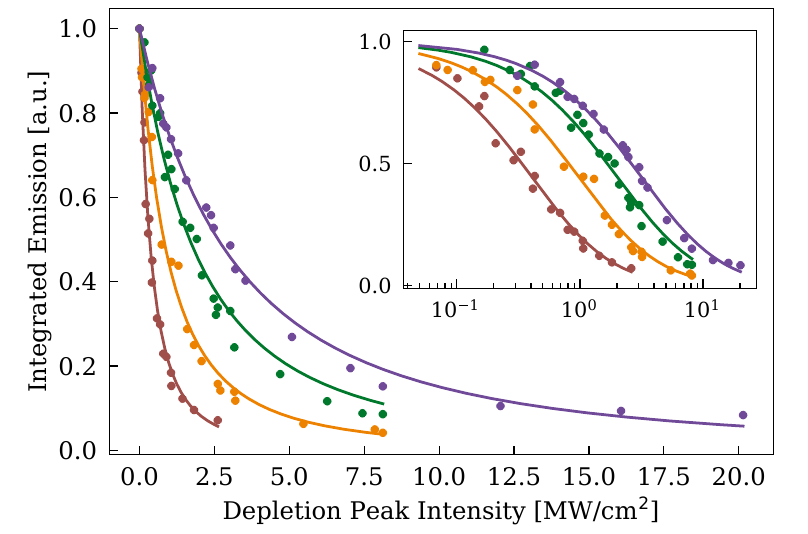}
  \caption{Single-nanoparticle integrated luminescence as function of the peak intensity of the 808-nm depletion beam. The experimental data of core-shell samples $92/8$, $88/12$, $84/16$, and $80/20$ are represented in dark red, orange, dark green, and purple, respectively. For a reliable comparison, the excitation intensity for each sample ($I_\textrm{exc} = 58$, $154$, $296$, and \SI{500}{\kilo\watt\per\centi\metre\squared}, respectively) was selected to deliver a luminescence yield of about \SI{20}{cts\per\milli\second}. The continuous curves correspond to the best fit of the function $(1 + I_\textrm{808} / \beta)^{-2}$, where the fitting parameter $\beta$ helps to estimate the characteristic depletion intensity $I_\textrm{depl} = (\sqrt{2}- 1) \cdot \beta$ (equal to \SI{0.34(6)}{}, \SI{0.8(1)}{}, \SI{1.7(3)}{}, and \SI{2.6(3)}{\mega\watt\per\centi\metre\squared}, respectively) at which the integrated emission yield is halved. Inset: Experimental data and curves represented in logarithmic scale.}
  \label{fgr:depl}
\end{figure}

Large variations in the lanthanide ratio and the coating of an inert shell may significantly alter the depletion efficiency induced by the 808-nm laser, which is critical for the optimal performance of STED microscopy. We studied the integrated luminescence of a single nanoparticle as function of the peak depletion intensity of the $808$-nm Gaussian beam.
To effectively compare the four types of nanoparticles, each sample was imaged at excitation intensities sufficient to deliver a luminescence yield of about \SI{20}{cts\per\milli\second} in the absence of the depletion laser. Figure \ref{fgr:depl} illustrates the normalised depletion curves of the core-shell nanoparticles, where we can estimate the characteristic depletion intensity $I_\textrm{depl}$ at which the emission yield is halved. We observe that $I_\textrm{depl}$ shifts towards lower values when increasing the lanthanide ion ratio, in a similar manner than the excitation intensity $I_\textrm{exc}$. Interestingly, the ratio between the depletion and the excitation intensity $I_\textrm{depl}/I_\textrm{exc}$ for each sample assumes values between $5.2$ and $5.8$, demonstrating a comparable depletion efficiency across each of the four nanoparticle compositions.

\begin{figure}[!t]
  \centering
  \includegraphics{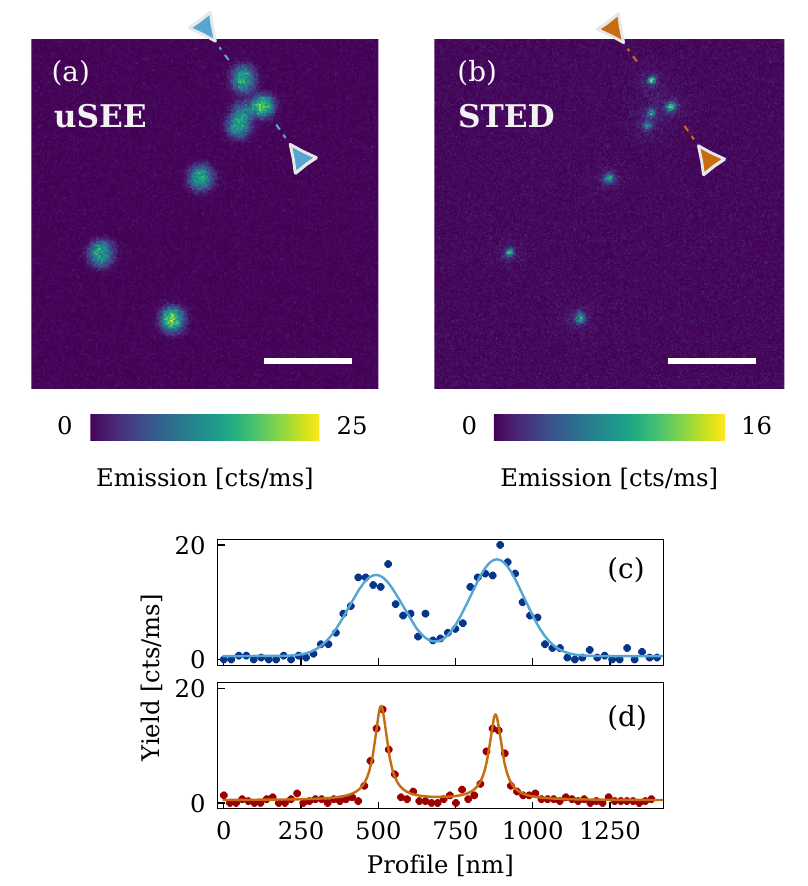}
  \caption{Super-resolution imaging with core-shell nanoparticles with $\eta = 92/8$ (NaYb$_{0.92}$Tm$_{0.08}$F$_4$@NaYF$_4$). (a) Confocal $x$-$y$ image of the UCNPs by means of the uSEE technique (peak intensity of the excitation beam: \SI{46}{\kilo\watt\per\centi\metre\squared}). Image size: $200 \times \SI{200}{pixels}$. (b) Same confocal imaging scan as in (a) performed with the STED technique. The peak power density of the 976-nm excitation CW laser and the 808-nm depletion beam are \SI{49}{\kilo\watt\per\centi\metre\squared} and \SI{4.1}{\mega\watt\per\centi\metre\squared}, respectively. Image size: $250 \times \SI{250}{pixels}$. Dwell time: \SI{3}{\milli\second}. Scale bar: \SI{1}{\micro\metre}. (c) Line profile of the uSEE-imaged nanoparticles. The Gaussian best-fit gives a FWHM of \SI{206(12)}{\nano\metre} (d) Line profile of the STED-imaged nanoparticles. The Lorentzian best-fit returns a FWHM of \SI{57(5)}{\nano\metre}. The relative heights of the emission peaks in STED mode compared to uSEE mode are affected by spatial sampling and image noise, as described in Supplementary Figure 6.}
  \label{fgr:super-res}
\end{figure}

The ability to shift the emission curve over a large power region, while maintaining strong non-linear behaviour, has a great impact on super-resolution imaging techniques such as uSEE and STED. The first advantage is to reduce significantly the laser power required to access these super-resolution regimes.
We confirmed this milestone by studying the core-shell nanoparticles with the highest $\eta$ (i.e.\ $92\%$ Yb, $8\%$ Tm), corresponding to the case with lowest intensity threshold. By imaging with only the $976$-\SI{}{\nano\metre} CW laser at a power density compatible with the super-linear emission region, we were able to access the uSEE regime and measure a single-nanoparticle resolution of about \SI{200}{\nano\metre} (Figure \ref{fgr:super-res}(a,c)). Resolutions of about half the diffraction limit by means of uSEE nanoscopy were already attained by our group \cite{Denkova2019}, but in this case we employed an excitation peak intensity as low as \SI{46}{\kilo\watt\per\centi\metre\squared} (about \SI{85}{\micro\watt}), nearly three times lower than previously reported. Additionally, we tested the efficiency of the STED technique by co-illuminating the sample with both the excitation $976$-\SI{}{\nano\metre} laser and the annular-shaped $808$-\SI{}{\nano\metre} depletion beam. In this case, we achieved single-nanoparticle resolution (Figure \ref{fgr:super-res}(b,d)), demonstrating that the depletion process in the Tm$^{3+}$ ions remains effective despite the high concentration values of the Yb$^{3+}$ ion. The best resolution observed is comparable to that of current state-of-art UCNP-based STED nanoscopy \cite{Peng2019,Liu2017}, though the peak intensity of the depletion beam was only \SI{4.1}{\mega\watt\per\centi\metre\squared} ($\SI{19}{\milli\watt}$), more than two times smaller than our previous report \cite{Liu2017}. We clarify that the excitation power for the STED technique is selected to minimise the radiation fluence delivered to the sample, while maintaining a sufficient signal-to-noise ratio. The excitation powers for STED could be lowered to values comparable to those used in uSEE, which is beneficial for improving both lateral and transversal resolutions \cite{Ploschner2020}.

We also analysed the optical properties of the core-shell nanoparticles with the highest slope value, i.e.\ $\eta = 84/16$ (as reported in Figure \ref{fgr:core-shell}), confirming similar resolution performance for STED and even better resolutions for uSEE (Supplementary Figure 7). 
The improvement observed in uSEE nanoscopy reflects well the steepening of the emission curve over the different concentration ratios. Resolution enhancement of non-linear emitters with slope $s$ can be practically estimated to be proportional to $\sqrt{s}$, giving a relative improvement of $13\%$ when increasing the slope from $5.6$ to $7.2$ (corresponding to a change in $\eta$ from $92/8$ to $84/16$). This is in good agreement with the measured lateral resolution, improving from $206$ to \SI{183}{\nano\metre} (Figure \ref{fgr:super-res}(c) and Supplementary Figure 7(c)).
Moreover, the reduced excitation powers required for the core-shell samples can be clearly observed by comparing these results with the performance of the corresponding core-only nanoparticles (see Supplementary Figure 8).

It is evident that the ability to achieve super-resolution operation at different lanthanide concentrations ($\eta$ varying from $92/8$ to $84/16$) largely relaxes the constraints on the crystal composition. Conventional UCNP synthesis for STED and uSEE applications has required specific amounts of the matrix ions and the lanthanide dopants, i.e.\ 20\% Yb and 8\% Tm. Instead, the design of the fully lanthanide-based nanocrystal presented in this work is simply based on the ratio between sensitiser and activator, alleviating requirements on precise concentration values during the UCNP synthesis.

UCNP-based microscopy offers several advantages in terms of long-term photostability and near-infrared working conditions, including significant benefits such as low autofluorescence background, small scattering/absorption, and low photodamage. On the other hand, the limited photon budget of the two super-resolution techniques requires an image acquisition time in the range of $60$-\SI{180}{\second}, a few orders of magnitude slower than fluorescence super-resolution methods \cite{Sahl2017,Sigal2018}. However, combination with other imaging techniques can substantially improve the current performance. For instance, the development of parallelised imaging systems with multiple beams scanning sub-areas of the sample has enabled a manyfold increase in the frame rate \cite{Bingen2011}. In addition, non-linear SIM \cite{Ingerman2019} is compatible with the super-linear emission feature of uSEE microscopy, representing a valid solution to enhance both imaging speed and resolution performance.

\begin{figure}[!t]
  \centering
  \includegraphics{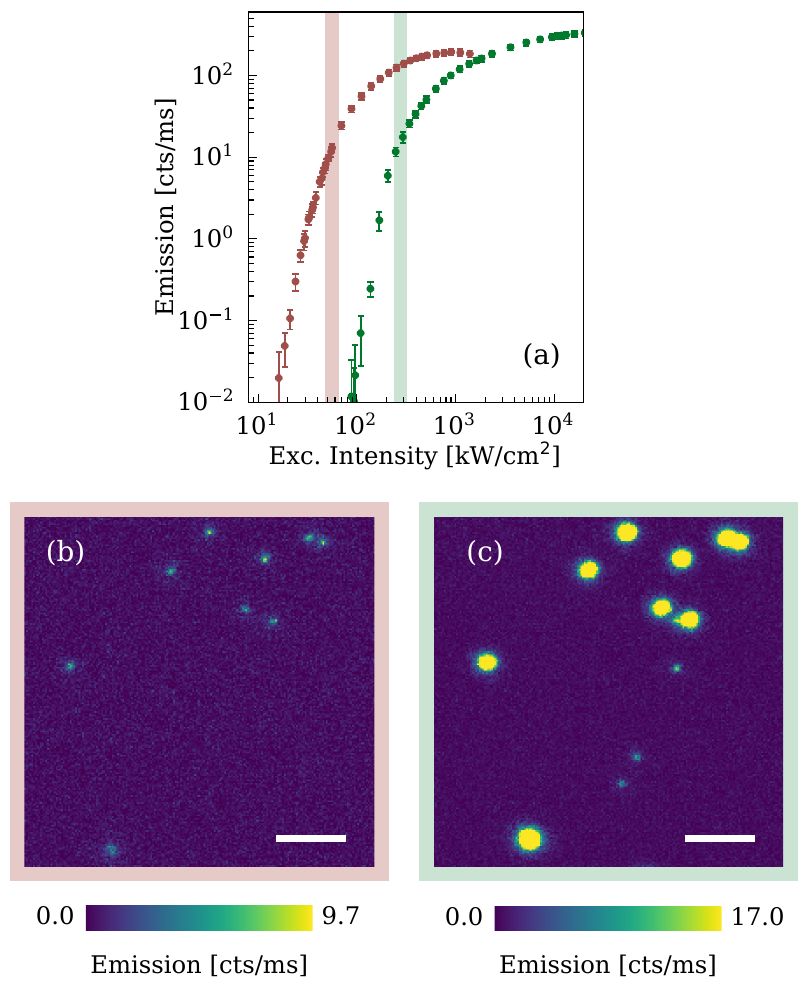}
  \caption{Multiplexed STED imaging of the core-shell UCNPs with $\eta = 92/8$ and $84/16$. (a) Single-nanoparticle $455$-\SI{}{\nano\metre} emission curve of each sample, represented in dark red and dark green, respectively. (b-c) STED confocal images of the two UCNP samples mixed and deposited on the same microscope slide. We can distinguish the first type in (b), excited at $I_{976} = \SI{53}{\kilo\watt\per\centi\metre\squared}$ (peak intensity of the depletion beam $I_{808} = \SI{4.3}{\mega\watt\per\centi\metre\squared}$). The second UCNP type is imaged in (c) at peak intensity $I_{976} = \SI{252}{\kilo\watt\per\centi\metre\squared}$ ($I_{808} = \SI{8.6}{\mega\watt\per\centi\metre\squared}$), while the former UCNP sample reaches the saturation regime. Image size: $200 \times \SI{200}{pixels}$. Dwell time: \SI{3}{\milli\second}. Scale bar: \SI{1}{\micro\metre}.}
  \label{fgr:multiplexed}
\end{figure}

Finally, controlling the emission curve of non-linear emitters allows more freedom in the adaptation of nanomaterials to each specific experiment and can lead to interesting applications. We took advantage of this effect to implement multiplexing capabilities in super-resolution imaging of UCNPs. Figure \ref{fgr:core-shell} shows that for large differences in lanthanide concentrations the non-linear emission regions of different nanoparticles can be distanced enough to avoid their possible overlap in excitation intensity. In Figure \ref{fgr:multiplexed}(a), we considered the case of the core-shell UCNPs with $\eta = 92/8$ and $84/16$, where the difference of the intensity threshold exceeds \SI{70}{\kilo\watt\per\centi\metre\squared}. By opportunely selecting the excitation power it is possible to perform super-resolution imaging of the first type of UCNPs (with no detectable blue luminescence from the second), or alternatively to acquire super-resolved images of the latter UCNPs while the former is subject to a saturated response. To experimentally verify this multiplexing technique, we deposited the two nanoparticle samples on the same microscope slide and acquired confocal STED images at peak power densities of the excitation beam equal to \SI{53}{\kilo\watt\per\centi\metre\squared} and \SI{252}{\kilo\watt\per\centi\metre\squared}. The two different UCNP samples can be clearly distinguished by comparing Figures \ref{fgr:multiplexed}(b) and (c). To visualise better the nanoparticles, the colour bar of the second image was rescaled to match the emission amplitude of the second type of particles with the higher intensity threshold, causing the first UCNPs to appear oversaturated. Inevitably the oversaturation imposes a blind region for the second type of nanoparticles, but in-plane this restriction is limited only to an area with radius smaller than \SI{160}{\nano\metre}.
In different experimental conditions, nanoparticle aggregation may represent a limitation to the present multiplex imaging technique. Nevertheless, the two samples investigated here (i.e.\ $92/8$ and $84/16$) show a large shift in the emission curve, ensuring a luminescence difference of more than three orders or magnitude at the first control intensity (red strip in Figure \ref{fgr:multiplexed}(a)). Therefore, for small and medium aggregations ($\le 500$ nanoparticles), the luminescence comparison at the selected control intensities would be sufficient to discriminate the two nanoparticle types.

Shifting the upconversion luminescence to lower intensities is not only limited to STED nanoscopy. This straightforward multiplexing modality can also be applied to the uSEE technique (Supplementary Figure 9), as it involves the super-linear emission regime of UCNPs.

\section*{Conclusion}
The present work has investigated the single-nanoparticle emission of fully Yb-based UCNPs, by comparing different compositions and structures and scanning the excitation power over more than two orders of magnitude. We have confirmed that this class of nanoparticles, coated with a thin shell of inert (NaYF$_4$) material, effectively sustains the conditions for uSEE and STED super-resolution microscopy, despite the Yb/Tm concentration ratio varying from $80/20$ to $92/8$. In particular, the super-linear power-dependence of the $455$-\SI{}{\nano\metre} emission peak is characterised by slopes equal to or greater than $5.6$, with a peak value of $7.2$, ensuring in all cases a uSEE resolution at least twice better than the diffraction limit. Further, we have demonstrated that depletion efficiency of the $455$-nm emission remains effectively constant over a wide range of lanthanide concentrations, enabling STED resolution ($\sim \SI{60}{\nano\metre}$) to be largely unaffected. 

Increasing the Yb/Tm concentration ratio from $80/20$ to $92/8$ leads primarily to a shift of the emission curve towards lower excitation powers without affecting the non-linear optical properties of UCNPs, behaviour which opens several opportunities for super-resolution imaging. We have demonstrated that the laser power density required for uSEE and STED microscopy can be lowered by at least a factor of two. Additionally, we have performed multiplexed super-resolution imaging of a mixture of two different UCNP samples, purely through choice of different excitation depletion conditions for the first time.

These results will help future breakthroughs in UCNP-based bioprobes. From the perspective of nanomaterials, the simplified particle design introduces large flexibility in the concentration of the lanthanide ions, relaxing the constraints usually applied in the synthesis of UCNPs for imaging purposes. In the context of biological applications, imaging at lower excitation powers considerably reduces the risk of photo-damage in biological samples. Moreover, controlling the emission curve of the upconversion $455$-\SI{}{\nano\metre} signal enables the observation of two or more sub-cellular target structures with single-nanoparticle resolution, paving the way for relative location measurements and quantitative nanoscopy.

\section*{Experimental Methods}

\subsection*{\textnormal{\textbf{Materials:}}}
Ytterbium(III) chloride hexahydrate (YbCl$_3$, $99.99\%$), thulium(III) chloride hexahydrate (TmCl$_3$, $99.99\%$), yttrium(III) chloride hexahydrate (YCl$_3$, $99.99\%$), oleic acid (OA, $90\%$), 1-octadecene (ODE, $90\%$), sodium oleate ($\geq 82\%$, fatty acids), ammonium fluoride (NH$_4$F, $\geq 98\%$), and sodium hydroxide ($\geq 97\%$, pellets) were purchased from Sigma-Aldrich and used as received without further purification.

\subsection*{\textnormal{\textbf{Synthesis of Ln-based UCNPs:}}}
The Ln-based UCNPs were synthesised by means of a purpose-made automated growth system, implementing a modified co-precipitation method based on reported literature \cite{Liu2017,Shi2017}. The first step is the production of the optically active core of the nanoparticles NaYb$_x$Tm$_{1-x}$F$_4$, with $x = 0.92$, $0.88$, $0.84$, and $0.8$. A suitable ratio of YbCl$_3$ ($x$~\SI{}{\milli\mole}) and TmCl$_3$ ($1-x$~\SI{}{\milli\mole}) dissolved in methanol were stirred in a flask with OA (\SI{6}{\milli\litre}) and ODE (\SI{15}{\milli\litre}). The mixture was placed under argon gas flow and underwent a first heating process, consisting of \SI{30}{\minute} at \SI{75}{\celsius} and \SI{30}{\minute} at \SI{160}{\celsius}.
After a cooling down period, \SI{3.125}{\milli\mole} of sodium oleate and \SI{4}{\milli\mole} of NH$_4$F were added, and the resulting solution was stirred for \SI{30}{\minute} under argon gas.
The mixture was subject to a second heating process, \SI{160}{\celsius} for \SI{1}{\hour} and \SI{310}{\celsius} for \SI{40}{\minute}, and cooled down to room temperature under argon gas flow. The nanocrystals were isolated by dilution in ethanol and centrifugation, washed several times with ethanol/cyclohexane, and stored in cyclohexane.

A similar procedure is repeated a second time for the synthesis of the shell precursor NaYF$_4$, but starting with YCl$_3$ (\SI{1}{\milli\mole}) instead of the lanthanide elements. After the first cooling down period, the solution was stirred with NH$_4$F (\SI{4}{\milli\mole}) and sodium hydroxide (\SI{2.5}{\milli\mole}), instead of sodium oleate. Moreover, the second hearing process consisted of only a single step to \SI{160}{\celsius} kept for \SI{1}{\hour}.

For the nanoparticle coating, \SI{10}{\milli\litre} of ODE and an equal amount of OA were stirred in a flask together with \SI{3}{\milli\litre} of nanocrystals. The mixture was heated to \SI{75}{\celsius} under argon for \SI{30}{\minute}, and then further heated to \SI{310}{\celsius}. Subsequently, \SI{1}{\milli\litre} of as-prepared shell precursor was injected into the reaction flask and let react with the mixture for \SI{15}{\minute} at the same temperature. The same injection and waiting procedure was repeated several times. Finally, the products were cooled down to room temperature under argon gas exposure and purified with a similar procedure as for the core-only nanoparticles.

\subsection*{\textnormal{\textbf{Sample preparation:}}}
To prepare a sample slide, a cover slip (Grale HDS, HD LD2222 $1.01$P0, $22 \times \SI{22}{\milli\metre}$, No. 1) was washed with pure ethanol, followed by Milli-Q water, partially dried under nitrogen flow and left fully air-dry. \SI{50}{\micro\litre} of Poly-L-lysine solution ($0.1\%$ w/v in deionised H$_2$O) was dropped on the front surface of the cover slip and washed off with Milli-Q water after \SI{35}{\minute}. The UCNP solution was diluted with cyclohexane to a concentration of \SI{0.05}{\milli\gram\per\milli\litre}. \SI{20}{\micro\litre} of this solution was dropped onto the treated surface of the cover slip and immediately washed twice by \SI{500}{\micro\litre} of cyclohexane.
After being air-dried, \SI{10}{\micro\litre} of a index-matching medium is pipetted over a clean microscopy slide (Thermo Scientific, S41014A, $76 \times \SI{26}{\milli\metre}$) and the above prepared cover slip is gently placed on top. Slight pressure is applied to distribute the embedding medium. Finally, the slide sample was kept at room temperature to dry for at least \SI{48}{\hour}.

For the multiplexing measurement, the slide preparation followed a similar procedure. In this case, two UCNP samples were separalely diluted with cyclohexane to a concentration of \SI{0.05}{\milli\gram\per\milli\litre} and mixed together to form a $20$-\SI{}{\micro\litre} solution.

\subsection*{\textnormal{\textbf{Confocal system:}}}
A custom-made optical system developed by our group \cite{Denkova2019} (schematic illustrated in Figure \ref{fgr:set-up}) was employed to acquire the confocal images of UCNPs and measure their power-dependent emission curves.

The excitation beam is produced by a $976$-\SI{}{\nano\metre} CW laser diode (Thorlabs, BL976-PAG900). After collimation, the polarisation state of the beam is fixed by a linear polariser (Thorlabs, LPVIS100-MP). A pair of half-wave plate (Thorlabs, WPMH10M-980) and polarisation beam slitter (PBS, Thorlabs, CCM1-PBS252/M) is introduced along the optical path to control the laser power. The excitation beam power is varied by rotating the half-wave plate with a motorised rotation stage, and it is actively monitored by measuring the beam component reflected by the PBS. Moreover, a set of neutral density filters is employed to access lower ranges of the excitation power. 
The second laser beam required for the STED depletion is delivered by a $808$-\SI{}{\nano\metre} CW laser diode (Lumics, LU0808M250). After collimation, a PBS fixes the linear polarisation of the beam and a vortex plate (Holo/Or, VL-209-5-Y-A), precisely aligned along the optical path, modifies the beam profile into an annular shape. In this case, the depletion power is manually varied by changing the diode current.

The two NIR laser beams were spatially overlapped through a long-pass dichroic mirror (Chroma, ZT860lpxr), reflected by a second short-pass dichroic mirror (Chroma, T750spxrxt), and directed towards a quarter-wave plate (Thorlabs, AQWP10M-980) to produce the circular polarisation. Then, the laser radiation is focussed down onto the sample by a rigidly fixed microscope objective (Olympus, UPLSAPO 100XO, NA 1.4, 100x). Imaging of the beam profile on the focus point were performed for an accurate estimation of the intensity delivered to the sample (Supplementary Note 2). The sample slide is mounted on a piezo-electric 3-axis stage (Thorlabs, MAX311D/M with BPC203 piezo controller) to perform scanning acquisitions. The light emitted, scattered or reflected by the sample is collected back again by the objective, but only the upconversion signal is transmitted through the above mentioned short-pass dichroic mirror. In order to study the specific luminescence peak at \SI{455}{\nano\metre}, a narrow band-pass filter (Semrock FF01-448/20-25) selecting only the emission band at \SI{446(13)}{\nano\metre} is placed through the optical path. Finally, the signal is coupled into a multi-mode fibre (Thorlabs, M43L01, core diameter \SI{105}{\micro\metre}) and detected by a single-photon counting photodiode (Excelitas, SPCM-AQRH-14-FC). Precise synchronisation between the piezo-electric stage and the photodiode is realised by means of a multifunctional I/O card (NI 6353), which is able to deliver the required voltages for positioning the stage and to collect the digital counting signal from the detector.

\subsection*{\textnormal{\textbf{Acquisition procedure:}}}
A purpose-made LabVIEW program manages the entire scanning and acquisition procedure. Confocal 2D images are produced at fixed excitation/depletion powers, by scanning the sample and collecting the photon counts at each position (dwell time of \SI{3}{\milli\second} if not specified otherwise). The emission curves are analysed by keeping the stage at a fixed position and by acquiring the photon signal at different excitation powers. The automated control of the beam power reduces power uncertainties and minimises unnecessary laser exposures. Each point of the emission curve is obtained by acquiring the luminescence signal for \SI{1}{\second}, eight consecutive times. The values are averaged and the relative uncertainty is assigned with a confidence level of $95\%$. An additional uncertainty component, due to power fluctuations and estimated to change the luminescence signal of about $10\%$, is considered. Despite the dark count of the detector was \SI{e-1}{cts\per\milli\second}, the measurement sensitivity for the emission curve was lowered an order of magnitude by subtracting the background level. The depletion curves were obtained by acquiring mono-dispersed nanoparticle images at a fixed excitation power and varying the power of the depletion beam (with Gaussian profile shape). An averaged background component was subtracted to the integrated nanoparticle signal. Each depletion curve was normalised to the integrated luminescence yield with no depletion beam.

\section*{Conflicts of interest}
There are no conflicts to declare.

\section*{Acknowledgements}
This work has been supported by the Australian Research Council (ARC) funding through the Centre of Excellence for Nanoscale BioPhotonics (CE140100003) and by Macquarie University funding (Postgraduate Scholarship of Y.\ C., Research Fellowship of X.\ Z.). 
M.\ P.\ and Y.\ L.\ acknowledge support through Discovery Early Career Research Awards (DE170100241, DE170100821).
D.\ D.\ is grateful to the AGAUR (Generalitat de Catalunya) Beatriu de Pin\'{o}s grant (BP00269).


\bibliography{my_bib} 

\end{document}